\documentclass[aps,prb,amsmath,twocolumn,amssymb,titlepage,showpacs]{revtex4-1}

\usepackage{graphicx}
\usepackage{nicefrac}
\usepackage{amsfonts}

\usepackage{amssymb}
\usepackage{amsmath}

\usepackage{subfigure}
\usepackage{multirow}
\usepackage{tabularx}
\usepackage{array}

\usepackage{units}

\usepackage{tensor}
\usepackage{braket}

\usepackage{bm}
\usepackage{hyperref}

\usepackage{amssymb}
\usepackage{graphicx}
\usepackage{amsmath}
\usepackage{amsfonts}
\usepackage{color}

\begin{document}
\title{Symmetry Protected and Topologically Protected Crossed Andreev Reflection.}
\author{Yi-Ming Wu}
\affiliation{International Center for Quantum Materials, School of Physics, Peking University}
\affiliation{Collaborative Innovation Center of Quantum Matter, Beijing 100871, China}
\author{Xiong-Jun Liu}
\affiliation{International Center for Quantum Materials, School of Physics, Peking University}
\affiliation{Collaborative Innovation Center of Quantum Matter, Beijing 100871, China}

\begin{abstract}
Crossed Andreev reflection (CAR) is considered to probe the non-locality of the Majorana quasiparticles in topological superconductors. We propose here to study the CAR in multiband one-dimensional (1D) topological superconductors which may or may not have chiral symmetries, and show how the CAR can be applied to identify Majorana modes in such systems. In particular, for a multiband 1D topological superconductor with approximate chiral symmetry, both the multiple Majorana and low-energy Andreev bound modes can drive the nearly zero-bias CAR, which is shown to be protected by the chiral symmetry and is stable against non-magnetic disorder scatterings. The emergence of symmetry protected CAR makes it hard to identify exotic Majorana fermions out of low-energy Andreev bound states. It is interesting that the magnetic disorders, which break chiral symmetry, can fully kill the non-locality of low-energy Andreev bound states, while cannot affect the Majorana modes, leading to the topologically protected CAR which is solely contributed by the Marjorana modes. The experimentally relevant systems, including the semiconductor nanowires and the Fe chains in the top of an $s$-wave superconductor, have been considered.
\end{abstract}
\maketitle
\section{Introduction}

The exotic non-Abelian statistical property of Majorana zero modes (MZMs)\cite{Ivanov,Stern,Stone} in condensed matter systems makes them be promising candidates to realize fault-tolerant topological quantum
computation\cite{Chetan,Bonderson,Jason,Das}, and hence has generated broad interests of research. One way to obtain these quasiparticles is to use 1D p-wave superconductor~\cite{Kitaev1,Sengupta} or 2D $p+ip$ superconductor\cite{Kopnin}. However, intrinsic p-wave superconductivity is not necessary because it is equivalent to obtain an effective p-wave superconducting system by combining a 2D topological insulator and an s-wave superconductor\cite{Fu1,Fu2,Linder}, or a spin-orbit coupled semiconductor and a s-wave superconductor with Zeeman field\cite{Sau,Jason2,Lutchyn,Lutchyn2,Tudor,Oreg}. To experimentally identify MZMs it is suggested to measure zero-bias peak(ZBP) in the differential tunneling conductance dI/dV at the interface with a normal lead\cite{KTLaw,Flensberg,Xiong-Jun,Lin,Xiong-Jun2}. Recently these ZBPs have been observed in semiconducting nanowire systems\cite{Deng,Mourik} and in ferromagnetic atomic chains on a Pb superconductor\cite{Stevan,Pawlak}. However for those realistic systems with multiple bands, the existence of chiral symmetry or approximate chiral symmetry leads to both multiple Majorana and low-energy and Andreev bound states(ABSs) in the boundary, which are generically difficult to be distinguished due to the small minigaps\citep{Sumanta,Yuqin}.

Alternatively, one can detect these exotic MZMs by observing crossed Andreev reflection(CAR) which reveals their non-locality property\cite{Nilsson,JieLiu}. This is usually achieved by measuring the shot noise at low temperature and low bias voltage. In our work we investigated 1D topological superconductors with and without approximate chiral symmetries and their transport properties which can reveal CAR effect. Both semiconducting nanowire model and Fe chain model are considered. For such 1D systems with approximate chiral symmetry, both MZMs and ABSs can give rise to the nearly zero-bias CAR, which is shown to be protected by the chiral symmetry and can be referred as symmetry protected CAR compared to the topologically protected one which only comes from MZMs. The emergence of symmetry protected CAR makes it hard to identify exotic MZMs out of low-energy ABSs.

Based on these considerations we then introduced two kinds of disorder, nonmagnetic and magnetic, into the system. Our results show that in the presence of nonmagnetic disorder effect, the non-locality for both MZMs and ABSs cannot be affected. As a result symmetry protected CAR still exists and impedes our way to identify really exotic MZMs. However in the presence of magnetic disorder, the non-locality of ABSs can be break down while MZMs remain unaffected. In this case only the MZMs can contribute to CAR due to the topological protection. Low energy ABSs are sensible to magnetic disorder and hence are decoupled and localized to one boundary of the 1D system. It is in this way that the MZMs are detected. Our finding is especially useful in some realistic 1D systems where chiral symmetry is presence or weakly broken and there are multiple subbands crossing Fermi level. For example, there is approximate chiral symmetry and thus both boundary ABSs and MZMs in Fe chain system\cite{Yuqin}, but the commonly existing magnetic disorder in Fe atoms makes CAR a reliable method to identify MZMs.

This paper is organized in this way. In Section II we will introduce our models in detail and the Green's function method which we apply to calculate the tunneling current. In Section III we demonstrate the impact of disorder effect on the non-locality of different kinds of low energy states. In Section IV we will show our numerical results of Fano factors at zero bias and how CAR can be used to identify MZMs.

\section{Model and method}
In this section we will briefly introduce the models we used and the formalism of Green's function in calculating shot noise of tunneling current. Typically we have focused on 1D systems which can support the emergence of exotic states, and constrained our model in weakly breaking chiral symmetry. Here we introduce two kinds of model, a semiconducting system and a Fe chain system. The former is more illuminable in understanding our basic idea, while the latter are more closed to a realistic system.
\subsection{Semiconducting nanowire}
A quasi 1D superconductor resulting from combination of a semiconducting nanowire with $N_x$, $N_y$ and $N_z$ sites in the x, y and z direction and superconducting proximity effect can be modeled as\cite{Andrew},
\begin{equation}\label{eq.H}
\begin{aligned}
H_{SN}=-&\mu\sum_{\textbf{r},\alpha}c_{\textbf{r},\alpha}^{\dagger}c_{\textbf{r},\alpha}-\sum_{\textbf{r},\textbf{d},\alpha}t_{\textbf{d}}c_{\textbf{r},\alpha}^{\dagger}c_{\textbf{r}+\textbf{d},\alpha}\\
+&\Delta\sum_{\textbf{r},\alpha}(c_{\textbf{r},\alpha}^{\dagger}c_{\textbf{r},-\alpha}^{\dagger}+h.c.)\\
-&i\sum_{\textbf{r},\textbf{d},\alpha,\beta}U_{R,\textbf{d}}c_{\textbf{r},\alpha}^{\dagger}\hat{z}\cdot (\vec{\sigma}\times\textbf{d})_{\alpha\beta} c_{j,\beta}\\
+&\sum_{\textbf{r},\alpha,\beta}c_{\textbf{r},\alpha}^{\dagger}(V_z\sigma_z)_{\alpha\beta}c_{\textbf{r},\beta}.
\end{aligned}
\end{equation}
Here $\mu$ is chemical potential. $\alpha,\beta$ are spin index. \textbf{r} is the position of each site, and \textbf{d} is the vector from each site to its neighbors. We only consider nearest neighbor hopping, so here \textbf{d} denotes $\vec{d_x}$, $\vec{d_y}$ and $\vec{d_z}$ which are connect nearest neighbor sites in three directions. $\Delta$ is an s-wave superconducting pairing parameter. The Rashba interaction amplitude $U_{R,\textbf{d}}$ and hopping amplitude $t_{\textbf{d}}$ depend on the direction, which is based on the consideration of the anisotropy between different directions. To obtain an effective spinless system, a spin splitting potential $V_z$ is added in.

It can be found that without disorder, the transverse Rashba $(U_{R,\vec{d_y}})$ determines the symmetry class of the system. The presence of  $U_{R,\vec{d_y}}$ term in the Hamiltonian breaks the chiral symmetry of this system, so the symmetry class of this model is D class according to the ten-fold topological classification\cite{Alexander,Kitaev}. However, in the absence of the transverse Rashba term, the system is in BDI class, which supports integer number of MZMs at the two ends of the 1-D wire. From such symmetry reduction,  when there are even number of subbands crossing the Fermi level, an original phase in BDI class will support even number of MZMs and the reduced phase falls into the trivial one in the D class. This is due to the pair-coupling of MZMs on each end of 1D system. On the other hand, when the number of subbands crossing the Fermi level is odd then the system can give rise to the topological non-trivial phase in D class. One then can use the topological index $\nu$ defined by
\begin{equation}\label{eq.index}
(-1)^\nu=sgn[Pf(A_{k=0})Pf(A_{k=\frac{\pi}{a}})]
\end{equation}
 to distinguish topological and trivial phase, where $A_k$ is the BdG Hamiltonian written in Majorana basis\cite{Dindex}. For a system with multiple subbands crossing Fermi level, if the symmetry breaking is weak, then there will be low energy ABSs, which are still non-local, and thus hard to distinguish from MZMs.

\subsection{Fe chain}
In order to testify our findings in a realistic systems, we also considered system comprised of a single chain of Fe atoms. This is motivated by recent experiments where Fe chains are grown on a facet of Pb substrate to obtain an effective topological superconducting system. For a Fe single chain which is along x direction, the Hamiltonian writes,
\begin{eqnarray}\label{Eq.HFe}
H_{\rm Fe}&=&-\mu\sum_{\phi, \alpha,\bold x}c^\dag_{\phi \alpha}(\bold x)c_{\phi \alpha}(\bold x) \nonumber\\
&&+\sum_{\phi,\phi',\alpha,\bold x\neq\bold x'}t_{\phi\phi'}(\bold x,\bold x')c^\dag_{\phi \alpha}(\bold x)c_{\phi' \alpha}(\bold x')\nonumber\\
&&+J\sum_{\phi,\alpha,\beta,\bold x}c^\dag_{\phi \alpha}(\bold x)\sigma^y_{\alpha\beta} c_{\phi \beta}(\bold x)\nonumber\\
&&+i\lambda_{\rm so}\sum_{\phi,\phi',\alpha,\beta,\bold x}c^\dag_{\phi \alpha}(\bold x)(\vec l_{\phi\phi'}\cdot\vec \sigma_{\alpha\beta})c_{\phi \beta}(\bold x)\nonumber\\
&&+it_{R}\sum_{\phi,\alpha,\beta,\bold x}[c^\dag_{\phi \alpha}(\bold x)\sigma^x_{\alpha\beta} c_{\phi \beta}(\bold x+a\hat e_z)+h.c.]\nonumber\\
&&+\sum_{\phi,\bold x}\left[\Delta_{Fe}(\bold x)c_{\phi\uparrow}(\bold x)c_{\phi\downarrow}(\bold x)+h.c.\right],
\end{eqnarray}
Here $\phi$ denotes five d-orbitals for Fe atoms, and $\alpha,\beta$ label spin. ${\bold x}$ is the position of each Fe atom. As usual, $\mu$ is the chemical potential and $t_{\phi,\phi'}$ is the hopping amplitude from orbital $\phi'$ to orbital $\phi$. This Hamiltonian also includes a Stoner-theory spin splitting term characterized by $J$, an on-site spin-orbit coupling term $\lambda_{so}$ and a Rashba spin-orbit coupling term $t_R$. In addition, a proximity induced superconducting paring term $\Delta_{Fe}$ has been included. For a realistic system, we choose the parameters as $J=2.7eV$, $\lambda_{so}=60meV$, $t_R=0.1eV$, and $\Delta_{Fe}=1meV$. The hopping amplitudes $t_{\phi,\phi'}$ can be calculated using Slater-Koster integral\cite{Slater}, with hopping parameters $V_{\sigma}=-0.6702eV$, $V_{\pi}=0.576eV$ and $V_{\delta}=-0.1445$ for $m=0$, $|m|=1$ and $|m|=2$ bands respectively\cite{Zhong}.

Without on-site spin orbit coupling, this system has chiral symmetry with $T^2=1$, and the presence of $\lambda_{SO}$ breaks such symmetry\cite{Yuqin}. Thus same as our semiconducting nanowire model above, this single Fe chain model, with proper magnitudes for each parameter which are close to realistic system, can be categorized as D class in the presence of on-site spin orbit coupling term. Without such term it is in BDI class. However, this kind of symmetry breaking is very weak in realistic systems, leading to the emergence of many low energy ABSs.

This Fe chain model, together with the semiconducting model above, serve as examples of weakly chiral symmetry breaking systems, which are main interests of our study.

\subsection{General formalism}
Crossed Andreev reflection is the evidence of the non-locality of states, which allows one electron(hole) entering the superconducting nanowire from one lead, and reflecting one hole(electron) to the other lead. Two spatially separate Majorana modes can constitute a conventional fermionic state which can be highly non-local, therefore one can observe CAR in a system supporting MZMs. Likewise, low energy ABSs, resulting from a symmetry breaking procedure, can also be non-local and contribute to CAR if the symmetry breaking is not intense. Without any non-local states, one can observe local Andreev reflection in superconducting systems, which allows the injecting and reflecting process occurring at only one lead. For both crossed and local Andreev reflection there will always be 2 electrons(holes) entering the superconducting nanowire at each time.

To numerically study CAR, we have our 1D superconducting system grounded and the two leads biased at the same voltage and set the system at zero temperature. In such condition, the tunneling current in each lead is dominant by shot noise, which is due to the discrete nature of charge\cite{Blanter20001}. The shot noise correlator $P_{ij}=\int_{-\infty}^{\infty}dt\bar{\delta I_i(0)\delta I_j(t)}$ tells us details about the current correlation between lead i and lead j and can be used to identify whether there is CAR between i and j. The ratio of the shot noise correlator to the average current is called Fano factor, which measures how many electrons(holes) are injected during a current pulse.

We use the method of Green's function to study the property of tunnelling current, which depends on the Fisher-Lee relation\cite{Fisher},
\begin{equation}
S_{ij}^{\alpha\beta}=-\delta_{ij}\delta_{\alpha\beta}+i[\Gamma_{i}^{\alpha}]^{1/2}G^{r}[\Gamma_{j}^{\beta}]^{1/2}
\end{equation}
where $i,j=1,2$ denoting the lead 1 or lead 2 and $\alpha,\beta=e,h$ denoting electrons and holes. $G^r$ is the retarded Green's function given by
\begin{equation}
G^r=[EI-H_c-(\Sigma_{1}^{e})^{r}-(\Sigma_{1}^{h})^{r}-(\Sigma_{2}^{e})^{r}-(\Sigma_{2}^{h})^{r}]^{-1}
\end{equation}
$H_c$ is the superconductor Hamiltonian. The Green's function depends on the injecting energy $E$ of particles through the term $ EI $ with $I$ being an identity matrix. $\Gamma_{i}^{\alpha}$ is the coupling width which is given by $ \Gamma_{i}^{\alpha}=i[(\Sigma_{i}^{\alpha})^{r}-(\Sigma_{i}^{\alpha})^{a}]$ and $(\Sigma_{i}^{\alpha})^{r}=(\Sigma_{i}^{\alpha})^{a\dagger}$ is the self-energy:
\begin{equation}
(\Sigma_{i}^{\alpha})^{r}=\tau_{i}^{\alpha\dagger}(g_{i}^{\alpha})^{r}\tau_{i}^{\alpha}
\end{equation}
Note $(g_{i}^{\alpha})^{r}$ is the surface retarded Green's function of $\alpha$ particles in lead i and can be obtained using a rapid iteration method\cite{Sancho}. $\tau_{i}^{\alpha}$ the coupling matrix between $\alpha$ particles in lead i and the 1D topological superconductor.
Having known the scattering matrix, we now can compute the tunnelling current and the shot noise correlator using the following general expressions\cite{Nilsson,MPA},
\begin{equation}
\begin{aligned}
\bar{I_i}=&\frac{e}{h}\int_{0}^{eV}dE Tr[(1-\mathcal{R}_{ii}^{ee}+\mathcal{R}_{ii}^{hh})],\\
P_{ij}=&\frac{e^2}{h}\int_{0}^{eV}dE Tr[\mathcal{P}_{ij}],
\end{aligned}
\end{equation}
where $\mathcal{P}_{ij}$ is given by,
\begin{equation}
\begin{aligned}
\mathcal{P}_{ij}=&\delta_{ij}\mathcal{R}_{ij}^{ee}+\delta_{ij}\mathcal{R}_{ij}^{hh}-\mathcal{R}_{ij}^{ee}\mathcal{R}_{ji}^{ee}-\mathcal{R}_{ij}^{hh}\mathcal{R}_{ji}^{hh}\\
+&\mathcal{R}_{ij}^{eh}\mathcal{R}_{ji}^{he}+\mathcal{R}_{ij}^{he}\mathcal{R}_{ji}^{eh}\\
\mathcal{R}_{ij}^{\alpha\beta}=&\sum_{k}S_{ik}^{\alpha e}(E)S_{jk}^{\beta e\dagger}(E)
\end{aligned}
\end{equation}
The Fano factor is defined as
\begin{equation}\label{eq.fano}
F_{ij}=\frac{P_{ij}}{e(\bar{I_i}+\bar{I_j})/2}.
\end{equation}
For example, $F_{11}$ physically measures how many electrons are injected into the 1-D topological superconductor through lead 1 during a single noise pulse.

\section{Disorder effect}
\begin{figure}[h]
\begin{center}
\includegraphics[scale=0.25]{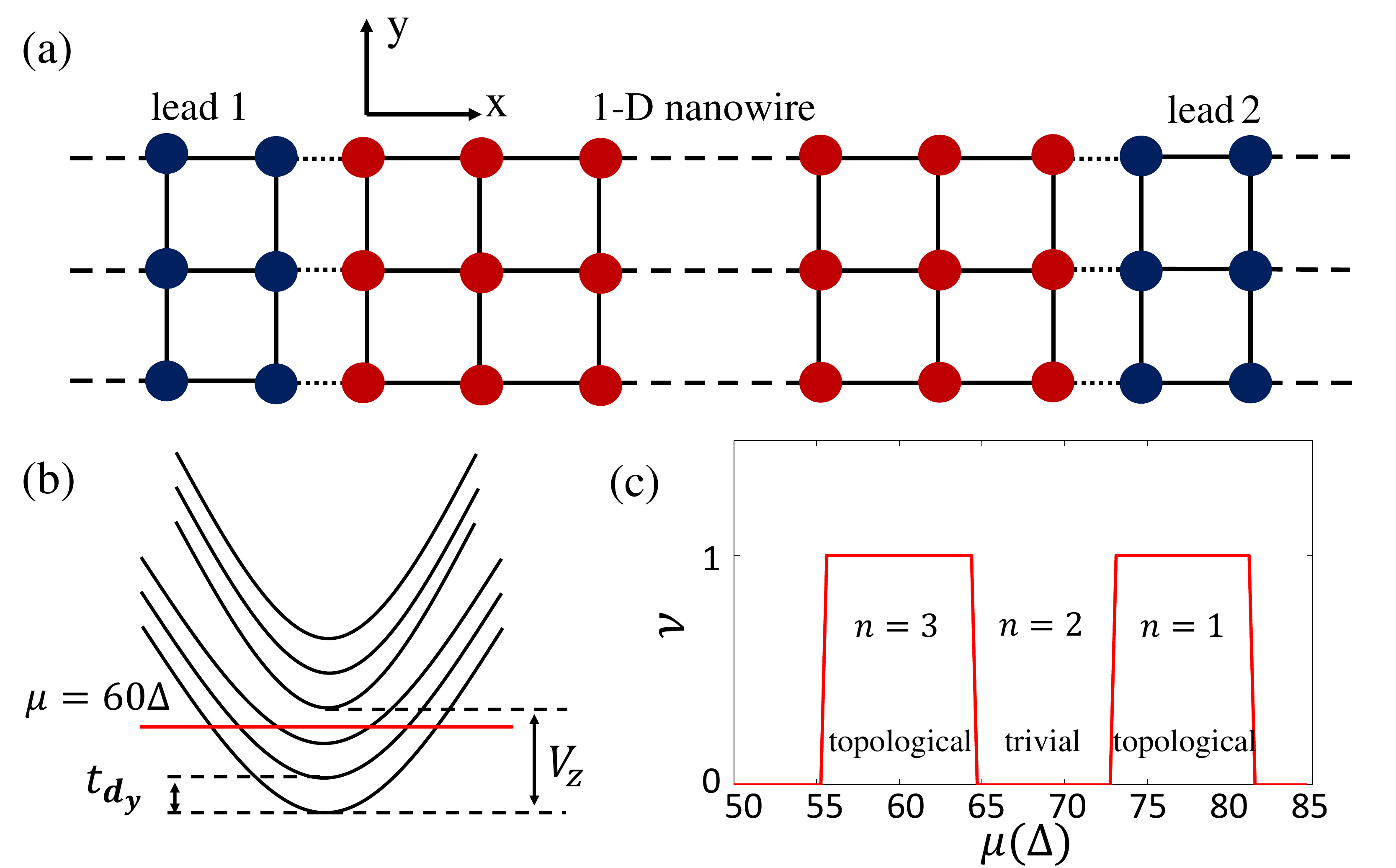}
\end{center}
\caption{(a).Model for calculation, with $N_y=3$, $N_z=1$ and $N_x$ remaining changable. Dasshed lines indicate the sites not shown, and dotted lines indicate the coupling between superconductor and normal leads. (b).Band structure of our model($\mu=60\Delta$) without superconducting gap. (c).Topological indices for the model. $\nu=1$ corresponds to topological phase, and $\nu=0$ corresponds to trivial phase. n is the numbers of subbands crossing the Fermi level.\label{Fig.1}}
\end{figure}

In this section we consider disorder effect on our models we have introduced. For the semiconducting nanowire system described by Eq.\ref{eq.H}, we take $N_y=3$ and $N_z=1$ to further simplify this model. This is depicted in Figure.\ref{Fig.1}(a), where a topological superconductor is coupled to two normal leads.
In our calculation, we choose the $\Delta=0.25meV$ as a unit\cite{Nadj}, and choose the other parameters as $ t_{\vec{d_x}}=5t_{\vec{d_y}}=30\Delta$, $ U_{R,\vec{d_x}}=10\Delta$, $V_z=13\Delta$. To obtain approximate chiral symmetry system and to study the effect of symmetry we choose $U_{R,\vec{d_y}}=0.1\Delta$. The chemical potential remains as a tunable parameter. A band structure plot of this model without superconducting gap is shown in Figure.\ref{Fig.1}(b). According to Eq.\ref{eq.index}, a plot of topological index for our weakly breaking chiral symmetry model is shown in Figure.\ref{Fig.1}(c) with varying chemical potential. The number of subbands crossing Fermi level is listed in the figure. For these parameters, the energies of ABSs are about $10^{-2}$ of superconducting gap, which is sufficiently low and thus can be treated as a weakly breaking symmetry system\cite{Sumanta}.

To take into account the disorder effect, we add both non-magnetic and magnetic disorder Hamiltonian to our models. Since non-locality is crucial for CAR, we are particularly interested in the influence of disorder on the non-locality of low energy states. To demonstrate this effect clearly, it is necessary to define the following quantity which can be called as intensity of non-locality,
\begin{equation}
\Theta=\frac{4P_{left}P_{right}}{(P_{left}+P_{right})^2}
\end{equation}
where $P_{left}$ and $P_{right}$ are the probabilities of finding a specific low energy state at the left side and right side of the 1D system. In our calculation the left side is defined as $n_p$ sites from left end to the middle of the wire, and so is the right side. Namely, we have
\begin{equation}
P_{left}=\sum_{x=1}^{n_p}|\psi(x)|^2, \quad P_{right}=\sum_{x=N_x-n_p+1}^{N_x}|\psi(x)|^2
\end{equation}
 $\Theta=1$ means the state is highly non-local, while $\Theta=0$ corresponds to a localized state. Details and results are shown in the following.

\subsection{Non-magnetic disorder}
\begin{figure}
\includegraphics[scale=0.4]{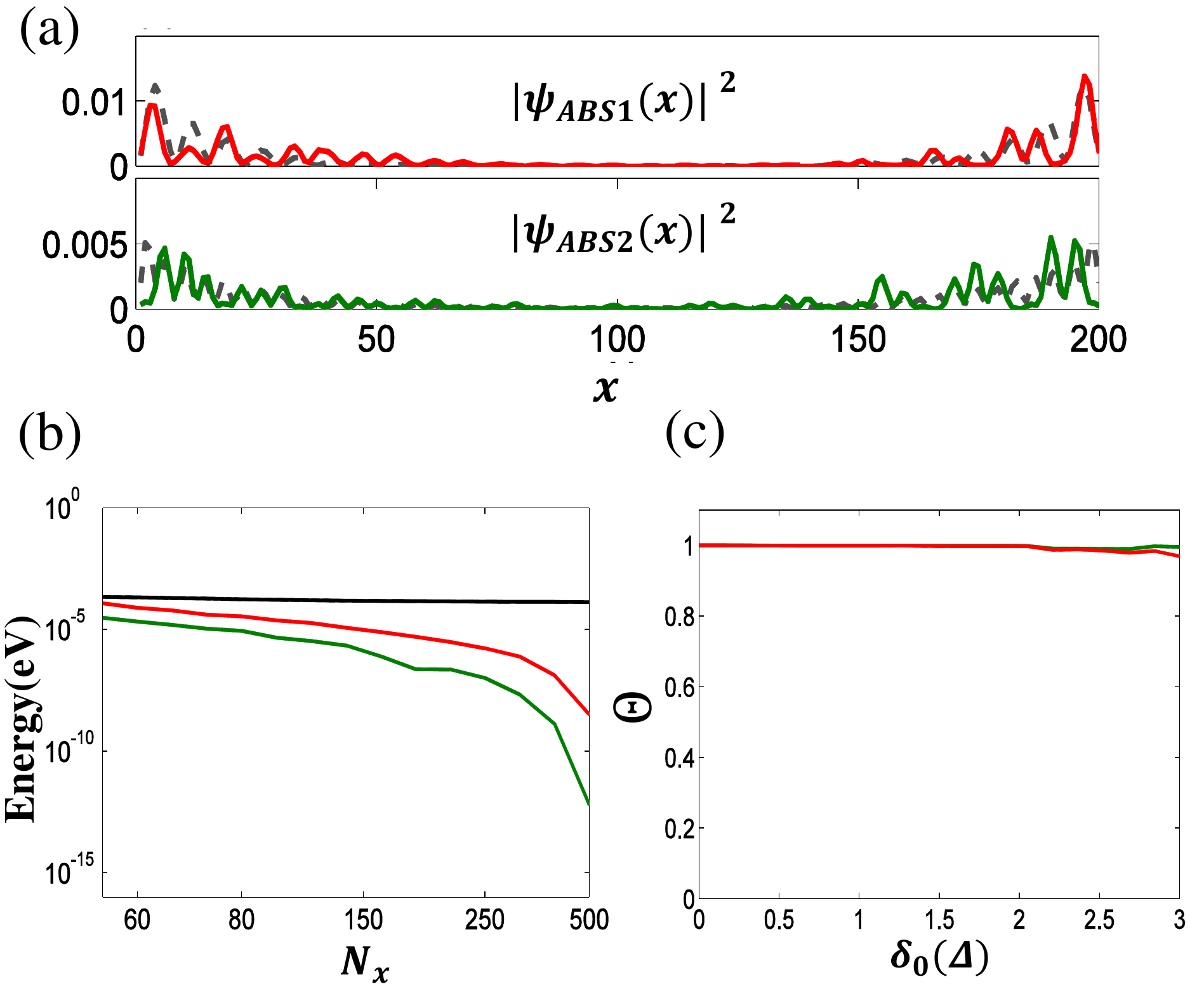}
\caption{(a)Impact of nonmagnetic disorder on two ABSs of 1D nanowire with $\mu=70\Delta$, $N_x=200$ and $\delta_0=3\Delta$. Gray dashed curves represent the wavefunctions of states without disorder. (b)Black curve represents gap energy of the system, while the two others represent the two ABSs energy for different wire length $N_x$. (c)Intensity of non-locality for the lowest two ABSs of $N_x=200$ varies with $n_p=40$ for different $\delta_0$.\label{Fig.3} }
\end{figure}

For our simplified 1D nanowire model, we include a non-magnetic disorder Hamiltonian in Eq.\ref{eq.H},
\begin{equation}
H_{nmd}=\sum_{\textbf{r},\alpha,\beta}c_{\textbf{r},\alpha}^{\dagger}(\delta_{\textbf{r}}\sigma_0)_{\alpha\beta}c_{\textbf{r},\beta}
\end{equation}
where $\sigma_0$ is a $2\times2$ identity matrix. The disorder potential $\delta_{\textbf{r}}$ is normally distributed on each site with a standard deviation $\delta_0$. Figure.\ref{Fig.3} shows the result of how non-magnetic disorder can affect the behavior of low energy states wavefunctions.
We choose $\mu=70\Delta$, for which the system is in trivial phase as indicated in the phase diagram of Figure.\ref{Fig.1}(c). There are two subbands crossing Fermi level, resulting two low energy ABSs. Finite size effect is worthy to notice here, because we need finite size effect to couple two ends of the wire to get CAR. Figure\ref{Fig.3}(b) shows the energy of the two ABSs with increasing $N_x$. As expected, a larger $N_x$ leads to a lower energy of ABSs, and hence a weaker finite size effect. To study the disorder effect we choose $N_x=200$. When the nonmagnetic disorder is turned on, the wavefunctions of these two ABSs are still distributed around the two end of the wire as shown in Figure\ref{Fig.3}(a), thus the presence of nonmagnetic disorder of $\delta_0=3\Delta$ is not able to cause localized ABSs. Different amplitudes of disorder are also considered. From figure\ref{Fig.3}(c) we can see the intensity of non-locality remains close to 1, revealing nearly no impact of disorder. If we keep increasing $\delta_0$ there is no evidence of localization for these states, and the superconductivity of the system will get destroyed.

\begin{figure}
\includegraphics[scale=0.3]{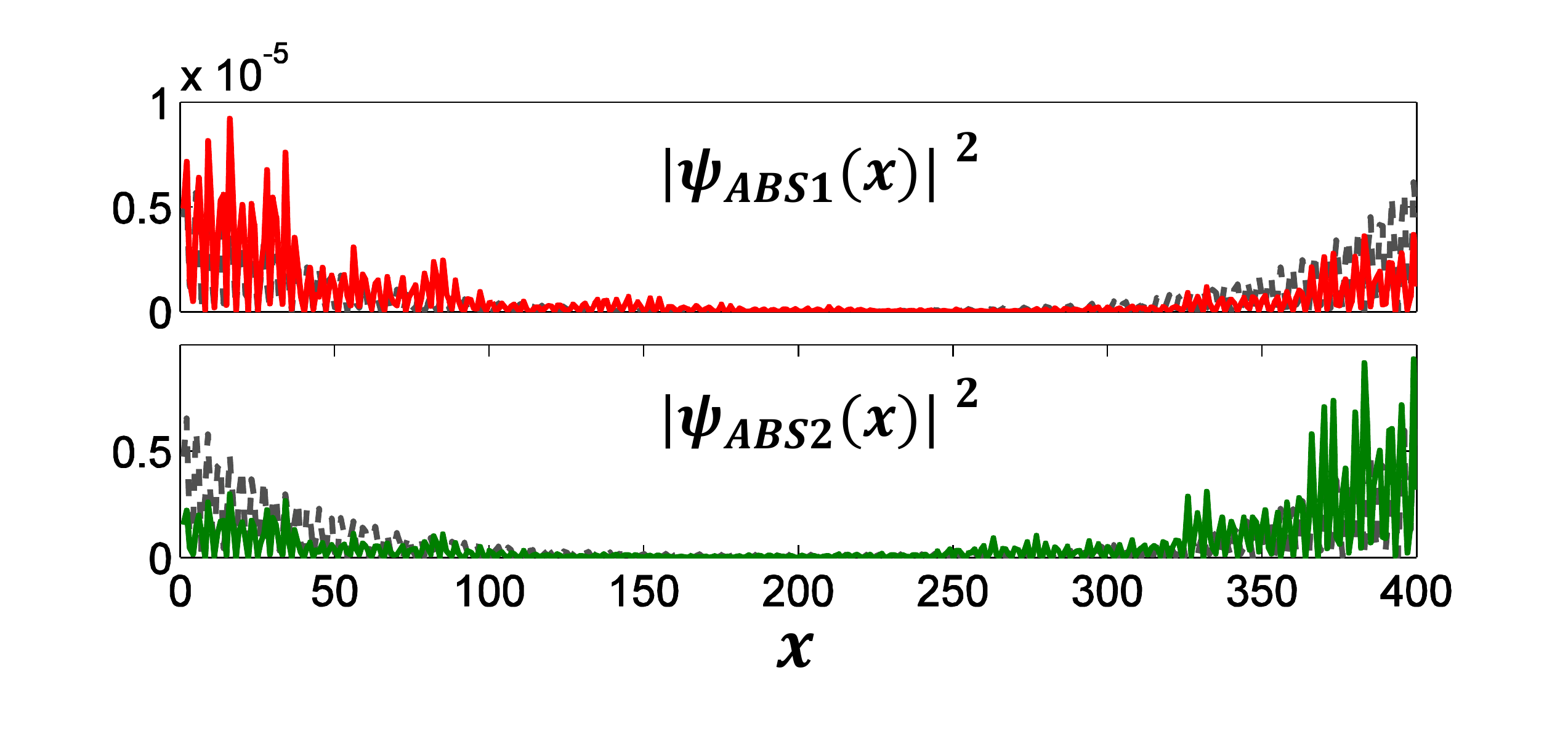}
\caption{Impact of nonmagnetic disorder on the lowest two Andreev bound states of a single Fe chain sytem with 400 atoms based on Eq.\ref{Eq.HFe}. The chemical potential is $\mu=2.1eV$,and the disorder amplitude is $\delta_0=7.5\Delta_{Fe}$. Other parameters in Eq.\ref{Eq.HFe} are kept the same as given in Sec.IIB. Gray curves correspond to states without disorder.\label{Fig.4} }
\end{figure}

For single Fe chain system, we add the following Hamiltonian to Eq.\ref{Eq.HFe},
\begin{equation}
H_{nmd}^{Fe}=\sum_{\phi,\alpha,\beta,\bold x}c^\dag_{\phi \alpha}(\bold x)(\delta_{\bold x} \sigma_0)_{\alpha \beta} c_{\phi \beta}(\bold x)
\end{equation}
where $\delta_{\bold x}$ is a normal distribution with standard deviation $\delta_0$. Here we choose $\delta_0=7.5\Delta_{Fe}$, and the chemical potential $\mu=2.1eV$. Other parameters are the same as given in Sec.IIB.
Figure.\ref{Fig.4} shows the wavefunctions of two low energy ABSs for a single Fe chain of 400 atoms in x direction. Similar to nanowire model, nonmagnetic disorder cannot break the non-locality of these ABSs. A higher value of $\delta_0$ will eventually break down the superconductivity but not the non-locality.
\subsection{Magnetic disorder}
So far we have only consider the effect of nonmagnetic disorder, we now show that a magnetic disorder can lead to localized ABSs.

For nanowire model, the magnetic disorder Hamiltonian is,
\begin{equation}
H_{md}=\sum_{\textbf{r},\alpha,\beta}c_{\textbf{r},\alpha}^{\dagger}(\delta_{\textbf{r}}\sigma_y)_{\alpha\beta}c_{\textbf{r},\beta}
\end{equation}
The magnetization of this disorder is in y direction. It easy to find that adding $H_{md}$ to Eq.\ref{eq.H} also breaks chiral symmetry, as transverse Rashba does.

\begin{figure}[h]
\includegraphics[scale=0.4]{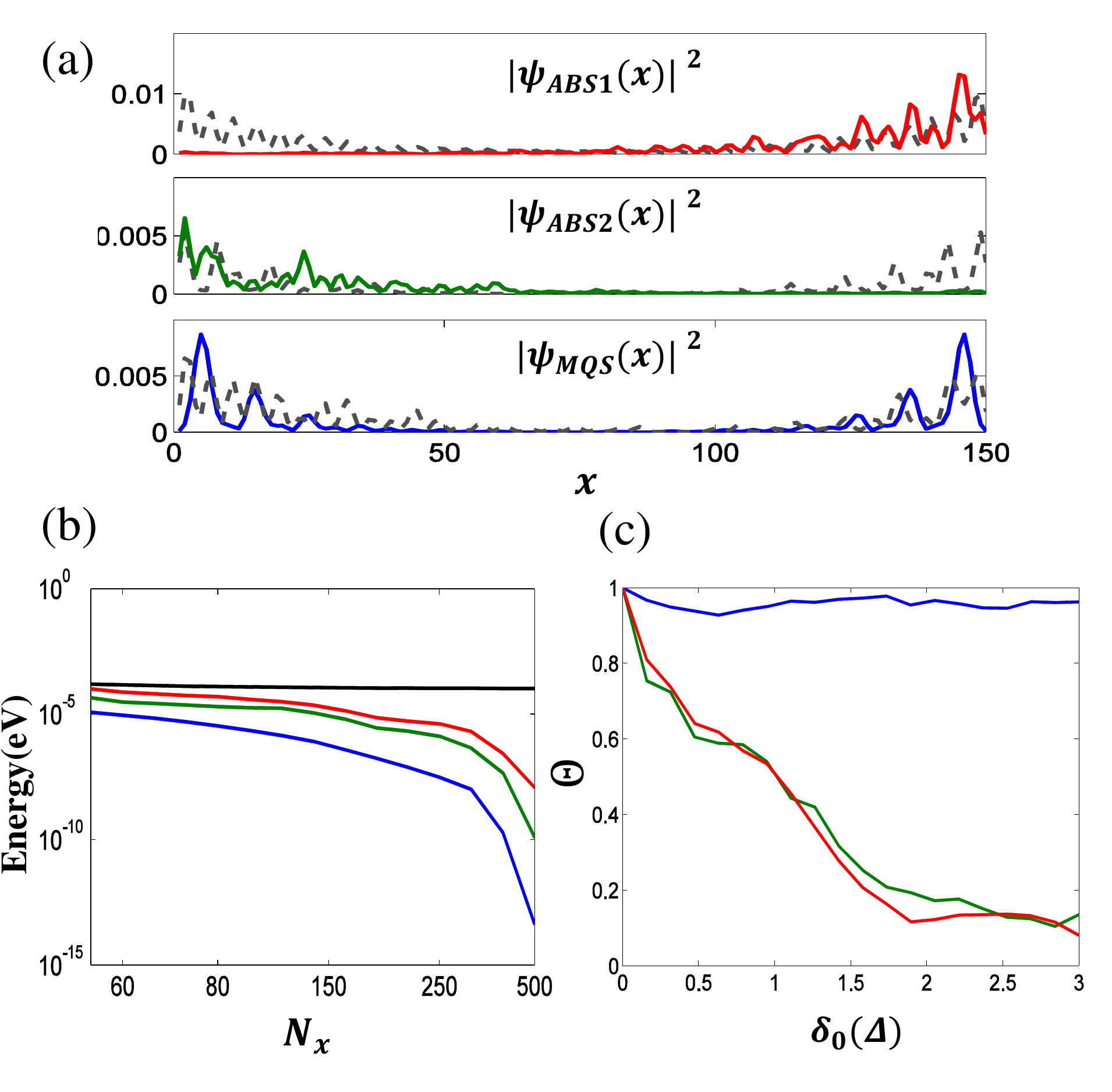}
\caption{\label{Fig.2}(a)Impact of magnetic disorder on three low energy states of 1D nanowire with $\mu=63\Delta$, $N_x=150$ and $\delta_0=3\Delta$. Gray dashed curves represent the wavefunctions of states without disorder. The red and green curves correspond to ABSs, and the blue one correspond to MZM. (b)Energy with different wire length $N_x$. THe black curve represents gap energy of the system, while the red and green ones represent the two ABSs and the blue one represent MZM. (c)Intensity of non-locality for the lowest MZM and two ABSs of $N_x=200$ varies with $n_p=30$ for different $\delta_0$. }
\end{figure}

Figure\ref{Fig.2} shows the magnetic disorder impacts for a topological phase with $\mu=63\Delta$ and $N_x=150$. There are three subbands crossing Fermi level, consequently there are two low energy ABSs and one MZM. When there is no disorder effect, all the boundary states are non-local under weak chiral symmetry breaking, as can be seen from these black dashed curves in Figure.\ref{Fig.2}(a). The magnetic disorder of $\delta_0=3\Delta$ then makes the low energy ABSs localized at one end of the 1D superconducting nanowire, as indicated by the red and green curves in the upper two subfigures. Since these states are not non-local anymore they cannot contribute to the CAR. This is true only when the finite size effect is not so strong that reasonable disorder can decouple the those states. However, due to the topological protection of the system, the Majorana modes can still constitute a non-local fermionic MZM even when magnetic disorder exists. For a trivial phase with only low energy ABSs, there will not be any non-local states surviving from magnetic disorder effect. Thus the disorder effects naturally distinguish the trivial and non-trivial phase in D class, by resulting in local and non-local low energy states. For the trivial phase, all the low energy states are localized to one end of the wire, but for the non-trivial phase, there will always be a non-local MZM and maybe several other localized low energy states. Figure.\ref{Fig.2}(b) shows $\Theta$ for different $\delta_0$.

\begin{figure}
\includegraphics[scale=0.25]{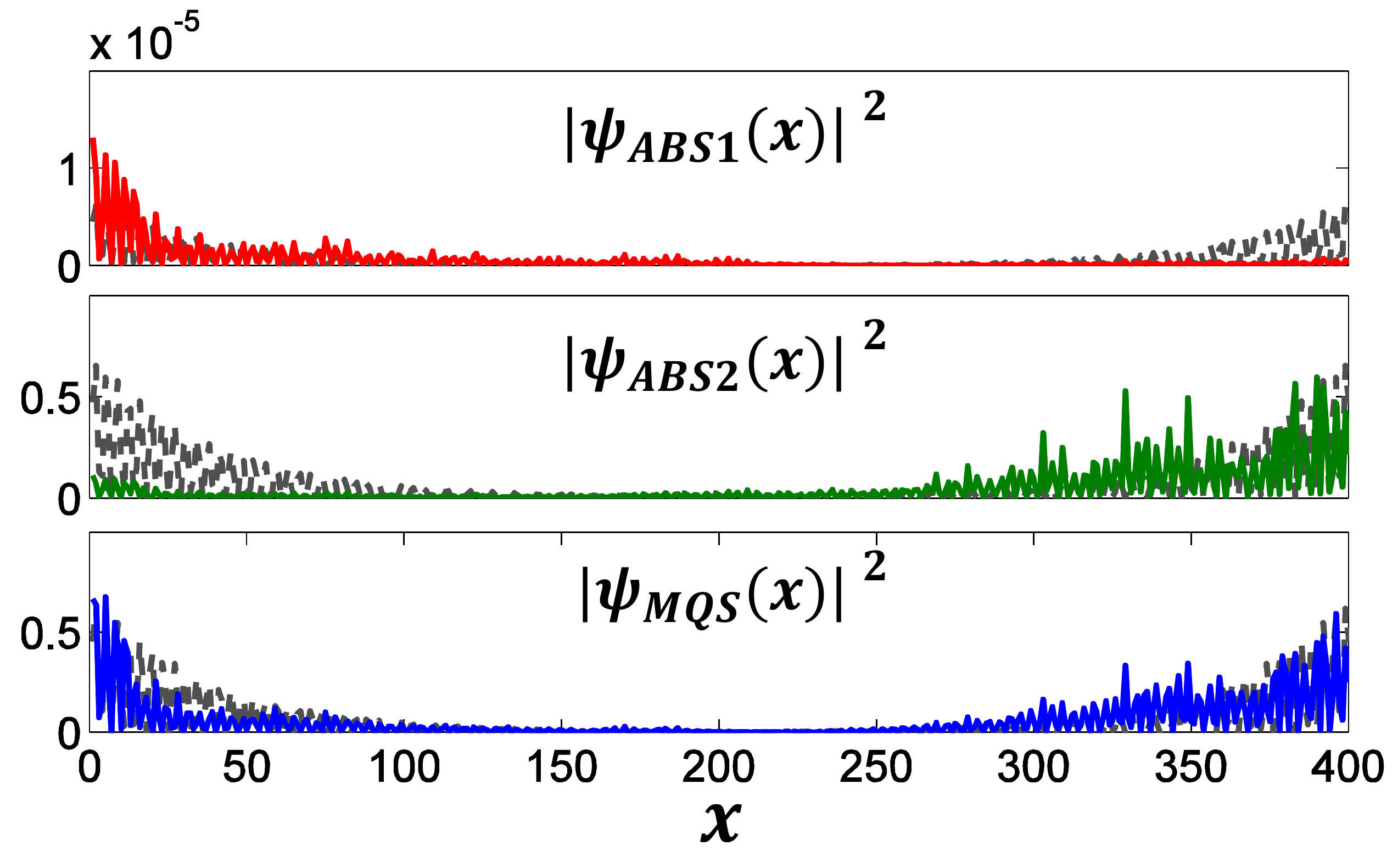}
\caption{Impact of magnetic disorder on the lowest three states of a single Fe chain sytem with 400 atoms based on Eq.\ref{Eq.HFe}. The chemical potential is $\mu=2.1eV$,and the disorder amplitude is $\delta_0=7.5\Delta_{Fe}$. Gray curves correspond to states without disorder.\label{Fig.5} }
\end{figure}

To solidify our results in real systems, we also studied the magnetic disorder effect on the Fe chain model, by adding the following Hamiltonian to Eq.\ref{Eq.HFe},
\begin{equation}
H_{md}^{Fe}=\sum_{\phi,\alpha, \beta,\bold x}c^\dag_{\phi \alpha}(\bold x)(\delta_{\bold x} \sigma_y)_{\alpha\beta} c_{\phi\beta}(\bold x)
\end{equation}
Here magnetic disorder couples $\sigma_y$ and can break chiral symmetry again. We have chosen the system in a topological phase ($\mu=2.1eV$) where there are odd number of bands crossing the Fermi level and made plot of wavefunctions for the lowest three states. From Figure.\ref{Fig.5} we can see the results are pretty much similar to those for our simple semiconducting nanowire model. Thus magnetic disorder which can break chiral symmetry can have a significant impact on low energy states for a given realistic system.

In order to observe CAR and use it as a probe to identify the existence of MZMs, one should expect there is reasonable finite size effect in the system. Because a very small finite size effect cannot strongly couple boundary states, there will not be correlated current between the two leads therefore one cannot observe CAR. For a system with reasonable finite size effect only magnetic disorder can leads to localized ABSs due to the symmetry reduction.

\section{Fano factor at zero bias}
In this section we study the disorder effect on CAR. As show in the previous section, disorder effect on our simple nanowire model and realistic Fe chain model share no difference, because both of them are weakly chiral symmetry broken. We therefore consider the nanowire model for simplification. Now the two leads in Figure.\ref{Fig.1} have to be taken into account. These leads are the same semiconducting system given by Eq.\ref{eq.H} but without superconducting paring term $\Delta$. The coupling strength between the normal leads and the superconducting nanowire is described by a hopping amplitude of $0.03t_{\vec{d_x}}$, where $t_{\vec{d_x}}$ is the hopping amplitude in x direction in Eq.\ref{eq.H}.

The emergence of CAR is due to the nonlocality of states. Unlike usual local Andreev reflection, CAR allows each of the two leads to tunnel only one electron during a current pulse, resulting a non-zero cross current correlator. When both the excitation energy and the coupling width between the leads and the 1-D topological superconductor nanowire are much smaller than the energy of the coupled Majorana modes, the local Andreev reflection can be suppressed in favor of CAR\cite{Nilsson}. We therefore consider Fano factors only at the zero bias, which is sufficient to reveal the basic picture. As a measurement of the CAR, the Fano factor of each lead($F_{11}$ or $F_{22}$) tells us how much electrons are tunneled from the lead to the superconducting nanowire during a current pulse. Therefore a Fano factor of 1 is a signal of crossed Adnreev reflection. But for a system with weakly breaking chiral symmetry, it is difficult to tell whether this signal is from MZM because there will be symmetry protected CAR. On the other hand, a Fano factor $F_{11}$($F_{22}$)$=2$ corresponds to local Andreev reflection consistent with the fact that there are two electrons entering the superconducting nanowire from lead 1(2). The crossed Fano factor $F_{12}=F_{21}$ measures the correlation between the two leads, therefore in local Andreev reflection where there is no correlation between two leads, $F_{12}$ should reach 0. In the following three different kinds Fano factor results based on Eq.\ref{eq.fano} are given for our weakly breaking chiral symmetry model.

\begin{figure}[h]
\includegraphics[scale=0.23]{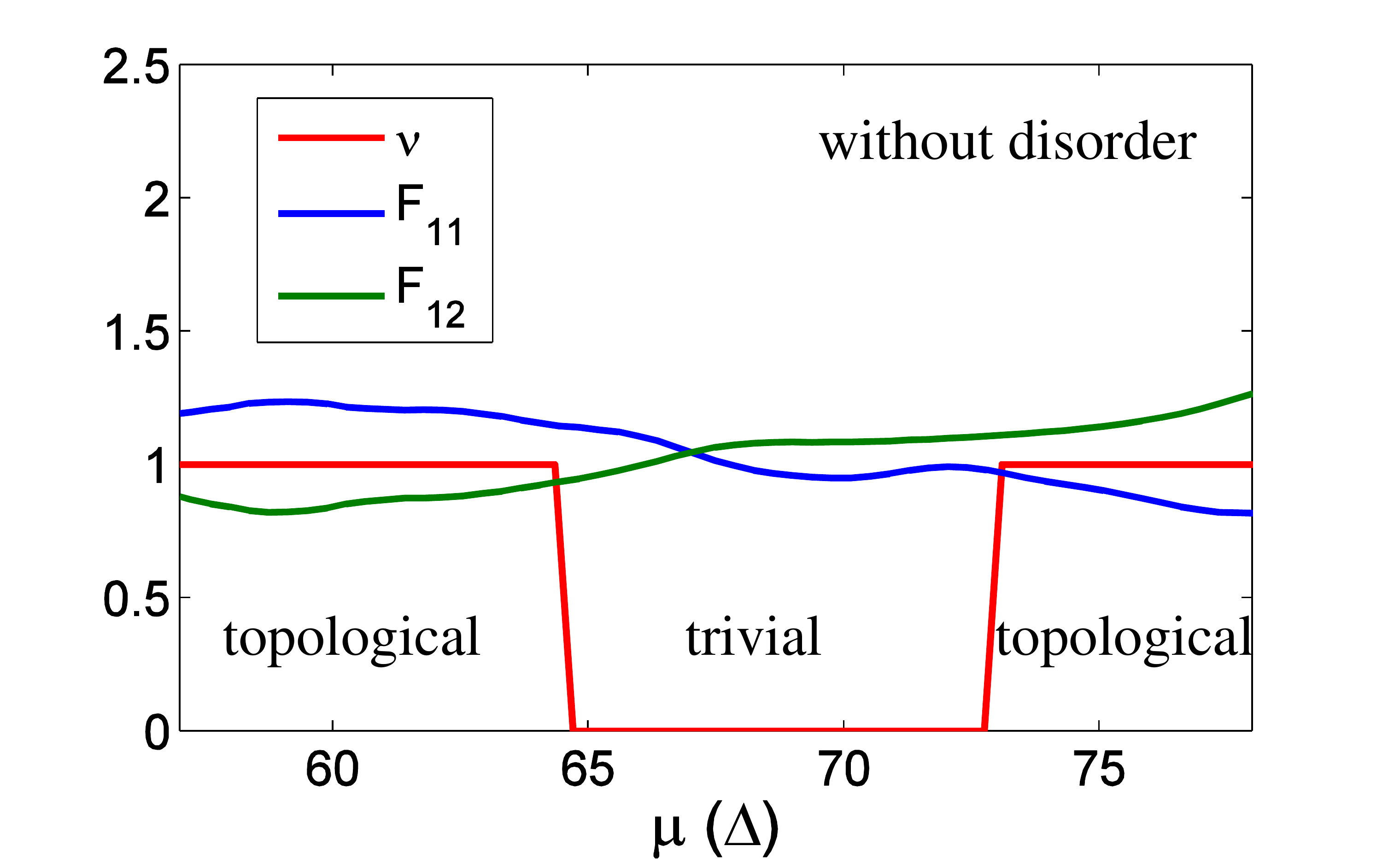}
\caption{\label{fano1} Topological indices and Fano factors at zero bias for different chemical potential without disorder effect. The nanowire length is $N_x=150$.}
\end{figure}
First of all, in the absence of disorder effect, all the low energy bound states are non-local so that even there are only ABSs one can still observe CAR. This kind of CAR can be referred as symmetry protected CAR. Our numerical results shown in Figure.\ref{fano1} confirmed this point. Both $F_{11}$ and $F_{12}$ are close to 1 because the tunneling current are dominates by correlated current between two leads. Note that these values are not exact 1 because the chiral symmetry breaking is very weak. If we increase the value of $U_R(\hat{y})$ to fully break the symmetry then both $F_{11}$ and $F_{12}$ are exact 1. Because these Fano factors are all close to 1 and don't change significantly with different chemical potential, we can hardly identify whether the system is in topological or trivial phase.
\begin{figure}[h]
\includegraphics[scale=0.23]{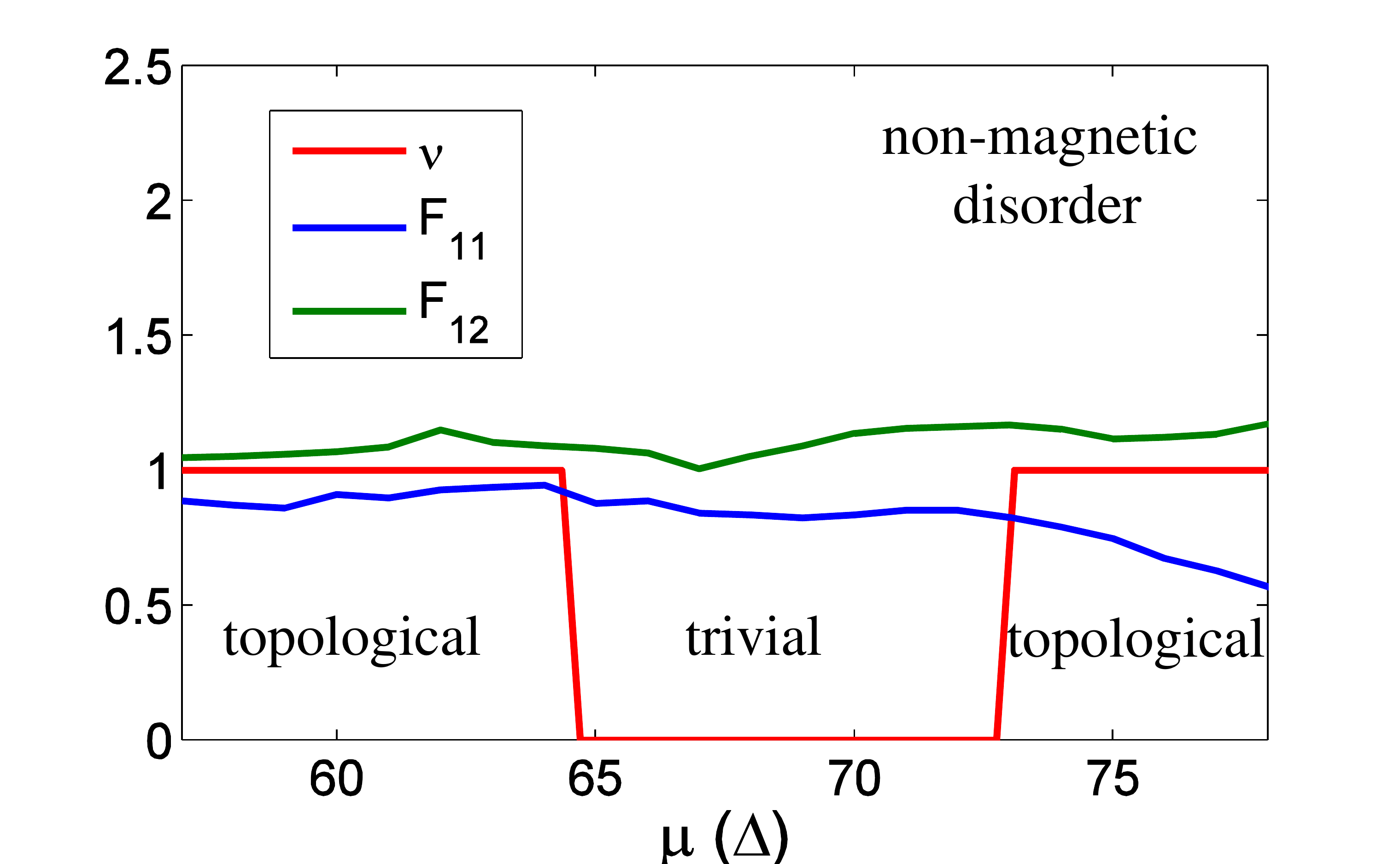}
\caption{\label{fano2}Topological indices and Fano factors at zero bias for different chemical potential with nonmagnetic disorder effect. The disorder is normally distributed on each site with a standard deviation $\delta_0=3\Delta$. The nanowire length is $N_x=150$.}
\end{figure}

Next a nonmagnetic normally distributed disorder with $\delta_0=3\Delta$ is applied in the system. As discussed above, since this kind of disorder cannot break chiral symmetry, one could expect there will be symmetry protected CAR in both topological and trivial phases. This is confirmed by our simulation results, shown in Figure\ref{fano2}. When chemical potential varies from about $65\Delta$ to $73\Delta$ the system is in trivial phase, but both $F_{11}$ and $F_{12}$ are close to 1 because of the symmetry protected CAR. Again we cannot discern the topological phases by measuring Fano factors.
\begin{figure}[h]
\includegraphics[scale=0.23]{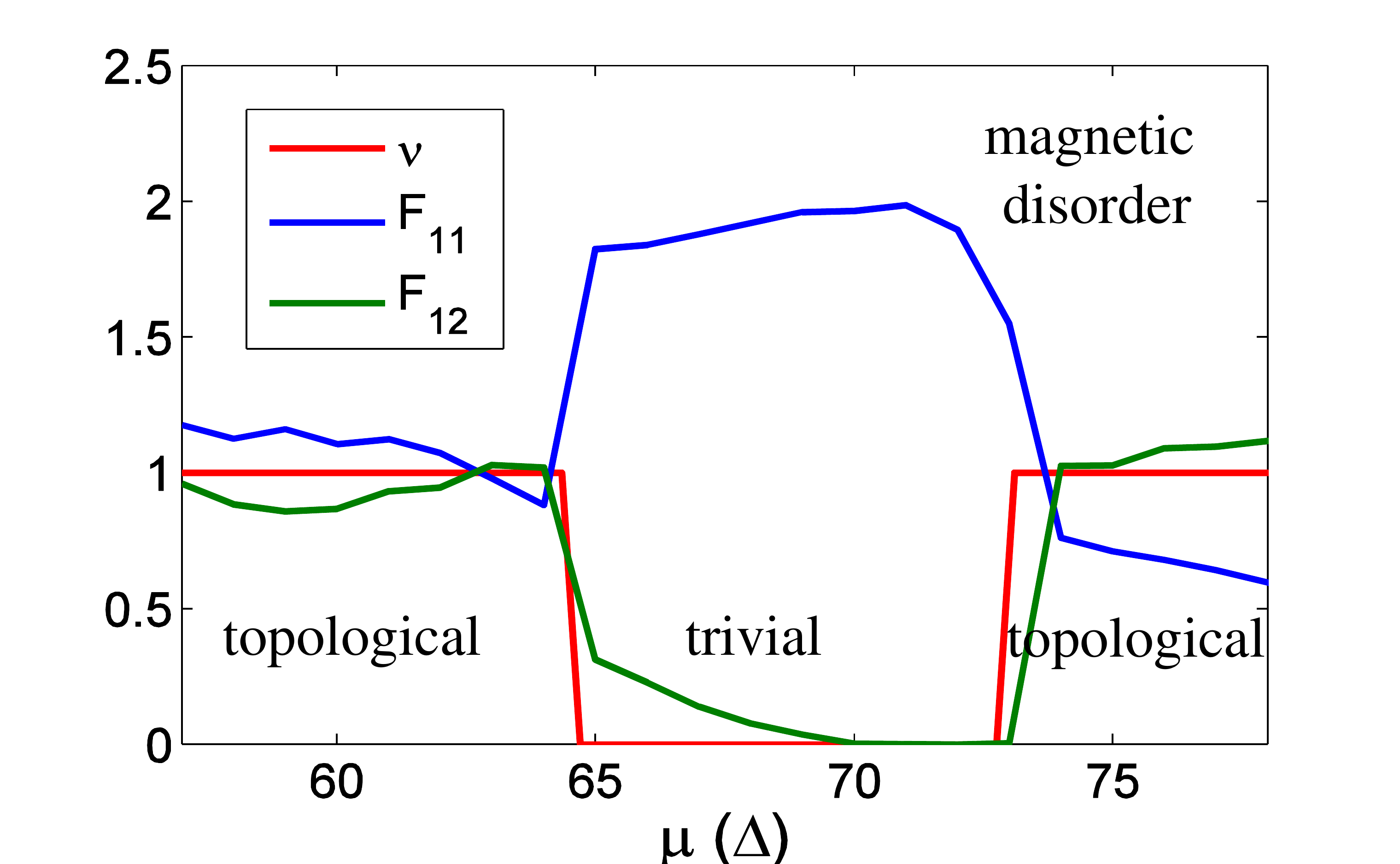}
\caption{\label{fano3}Topological indices and Fano factors at zero bias for different chemical potential with magnetic disorder effect. The disorder is normally distributed on each site with a standard deviation $\delta_0=3\Delta$. The nanowire length is $N_x=150$.}
\end{figure}

However introducing magnetic disorder which breaks the chiral symmetry yields a different story. Another calculation of Fano factors with magnetic disorder of $\delta_0=3\Delta$ gives the results in Figure.\ref{fano3}. From this figure we can see both $F_{11}$ and $F_{12}$ are close to 1 in the topological phase due to the topology protected CAR. In the trivial phase when there are even subbands crossing the Fermi level, $F_{11}$ reach around 2 and $F_{12}$ reach around 0, revealing there are only local Andreev reflections in this region. This is because our magnetic disorder breaks the symmetry and has localized these low energy ABSs resulting local Andreev reflection. Comparing these results we conclude that symmetry breaking is pivotal for identifying MZM in such 1D superconducting systems.

From these results we can see that, for a 1D system with weak chiral symmetry breaking, symmetry protected CAR can be switched to topology protected CAR. The latter can be used to identify MZM. Besides of the symmetry problem, there remains some issues worth notice. As mentioned above, the condition of the suppression of local Andreev reflection is weak coupling (between the leads and the nanowire) and the low energy tunneling. This requires that the coupling of the two Majorana boundary modes should not be too weak, which puts constraint on the up limit of the nanowire length. If the nanowire is very long with very small finite size effect, tunneling current may be contributed mainly from the local Andreev reflection. On the other hand, if the length of the nanowire is too small, then the strong finite size effect impedes the decoupling process of magnetic disorder. As a result, to use this method we would like the length of our nanowire to be proper, as too large or two small length will bring about difficulties in measurements.

\section{Conclusion}
In this study we find the requirement for identifying MZMs in a 1D topological superconductor based on non-local tunneling measurements: it is important to introduce magnetic disorder in order to overcome the ambiguity due to the presence of low energy ABSs. Magnetic disorder can break chiral symmetry and make low energy ABSs localized and thus kill the symmetry protected CAR, while MZMs are robust against such kind of disorder due to topological protection of the system. The topologically protected CAR may provide an unambiguous strategy to detect MZMs in topological superconductors.

\bibliography{crossed}

\end{document}